\documentclass[a4paper]{PoS}

\usepackage{subfigure}

\title{Searching
SUSY from below}

\ShortTitle{Searching
SUSY from below}

\author{\speaker{G. Grilli di Cortona}\\%
       SISSA - International School for Advanced Studies,\\
       Via Bonomea 265, I-34136 Trieste, Italy\\
       INFN - Sezione di Trieste,\\
       via Valerio 2, I-34127 Trieste, Italy\\
       E-mail: \email{ggrilli@sissa.it}}

\abstract{We studied the interplay between the mass reach for electroweakinos at future hadron colliders and direct detection experiments. The lack of new phenomena at the LCH motivates us to focus on split supersymmetry scenarios with different electroweakino spectra. A 100 TeV hadron collider may reach masses up to 3 TeV in models of anomaly mediation with long-lived thermal Winos. Moreover, in scenarios where the lightest neutralino is not the only dark matter component, the interplay between collider searches and direct detection experiments might cover large part of the parameter space.}

\FullConference{18th International Conference From the Planck Scale to the Electroweak Scale \\
		25-29 May 2015\\
		Ioannina, Greece }

\begin{document}

\section{Introduction}

After the discovery of the Higgs boson, the ATLAS and CMS experiments have constrained several beyond the Standard Model scenarios. In particular, in supersymmetric (SUSY) extensions of the Standard Model the lack of discovery of new coloured states put strong constraints on the spectrum, excluding squarks and gluinos with masses below about $1$ TeV. On the other hand, bounds on electroweakinos are much weaker, constraining Binos, Winos and higgsinos to be below about few hundred GeV. These constraints might be read as in tension with a natural implementation of supersymmetry: scenarios with heavy scalars and light electroweakinos favours split SUSY models. While giving up on the idea of solving the hierarchy problem, split SUSY ameliorates other potential problems such as FCNCs, CP violation and fast proton decay. It maintains, moreover, the successful unification of the gauge couplings and the lightest supersymmetric particle as a dark matter candidate. 

In SUSY models higgsino or Wino dark matter might have mass up to $\sim1$ or $\sim3$ TeV respectively, not directly accessible at the LHC. This motivates  the exploration of higher energies through 33 TeV or 100 TeV hadron colliders.  Therefore it is important to study the physics potential of future hadron colliders and to show the complementary roles collider searches and direct detection experiments can play. 

In Sect. \ref{sec:amsb} we describe the properties of the spectra in split SUSY models with anomaly mediation.  In Sect. \ref{sec:cr} we present the reach at future hadron colliders for split SUSY models with anomaly mediation with long-lived Winos, while in Sect. \ref{sec:dd} we show the interplay between direct detection and collider searches. Details for this scenario, reach at future colliders for other simplified models and the interplay between collider searches and direct detection experiments for split SUSY models with universal gaugino masses are described in the original work \cite{diCortona:2014yua}.

\section{Split SUSY with anomaly mediation}
\label{sec:amsb}

In split SUSY with anomaly mediation \cite{Randall:1998uk,Giudice:1998xp,ArkaniHamed:2004yi} the leading contribution to gauginos comes from one-loop anomaly mediation effects. However the Bino and Wino masses receive a further contribution from threshold effects:
\begin{equation}
M_{1,2}=\frac{\beta_{1,2}}{g_{1,2}}m_{3/2}+\frac{\alpha_{1,2}}{2 \pi}   \frac{(\tilde{m}^2+\mu^2)\mu\tan\beta}{(\tan^2\beta+1)\tilde{m}^2+\mu^2}\ln\left[ (1+\tan^{-2}\beta) \left(  1+\frac{\tilde{m}^2}{\mu^2}  \right)  \right],
\label{eq:AM}
\end{equation}
where $g_i$ is the corresponding gauge coupling, $\beta_i$ its beta function, $m_{3/2}$ is the gravitino mass, $\tilde{m}$ is the scalar mass-scale and the second term in the right hand side of the equation is the threshold effect. 
The model is entirely described by the gravitino mass $m_{3/2}$, the higgsino parameter $\mu$, $\tan\beta$ and the scalar mass-scale $\tilde{m}$. However $\tilde{m}$ and $\tan\beta$ are related by the value of the Higgs mass, bringing the number of free parameters to two once we fixed a value for $\tan\beta$.
Depending on the contribution of the higgsino, the nature of the dark matter candidate changes. Indeed for light higgsinos the higgsino itself is the LSP, with Wino NLSP. Winos can be the LSP if the higgsino is heavier than the Wino. The threshold corrections dominate when the higgsinos are very heavy and they lead to a Bino LSP. Bino dark matter, however, would be overproduced in the early universe and therefore it has to coannihilate with a Wino. Therefore the dark matter candidate can be a pure higgsino, a pure Wino, a mixed $\tilde{h}/\tilde{W}$ or a mixed $\tilde{B}/\tilde{W}$ state.   

\section{Reach at future hadron colliders}
\label{sec:cr}

\begin{figure} [t]
\begin{center} 
\includegraphics[scale=0.90] {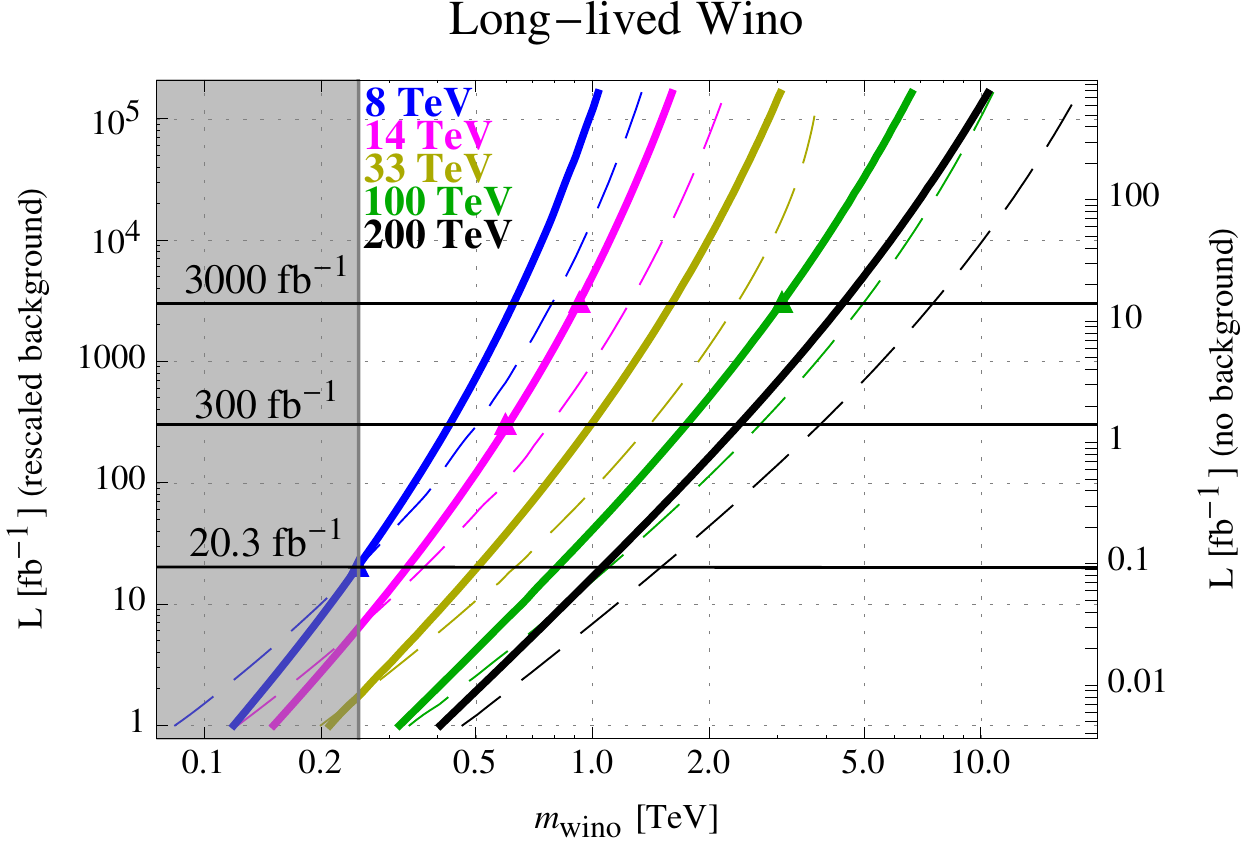}
\end{center}
\caption{Long-lived Wino. The left axis shows the integrated luminosity for the method explained in the text and on the right axis, the same for having $5$ events and no background. The grey shaded area is the current bound from \cite{Aad:2013yna}.} 
\label{fig:llWino} 
\end{figure}

Anomaly mediation models with Wino LSP can be probed by long-lived chargino searches. Heavy higgsinos can produce a spectrum in which the neutral Wino is highly degenerate with the charged Wino, while all the other states are decoupled. The mass splitting at tree level is suppressed and at one-loop is around 170 MeV \cite{Cheng:1998hc,Feng:1999fu,Gherghetta:1999sw}, implying that the charged Wino has a lifetime of order $c\tau\sim \mathcal{O}(10)$ cm.

The signature for long-lived Wino searches is one hard jet from initial state radiation, large missing energy and a disappearing track coming from the decay of the chargino to soft pions. The relevant background is due to unidentified leptons and charged particles with high mis-reconstructed transverse momentum as well as charged hadrons interacting with the inner detector. The ATLAS experiment already excludes charginos up to $250$ GeV in this scenario \cite{Aad:2013yna}.

The mass reach corresponding to an existing search at a new hadron collider can be obtained by computing the production cross section and requiring the same number of signal events needed to put the original bound \cite{CR,CR2}. We derived the mass reach rescaling the cut on the transverse momentum of the jet with the mass of the final state (solid lines in figure \ref{fig:llWino}) or keeping fixed the cut used by ATLAS in its analysis ($p_T>80$ GeV), if feasible (dashed lines in figure \ref{fig:llWino}). The right axis show the integrated luminosity requiring $5$ events and no background. 

This search is well motivated because a pure Wino is expected to thermally saturate the relic abundance around 3 TeV. Figure \ref{fig:llWino} shows the luminosity as a function of the mass reach for different hadron collider's centre of mass energy. LHC at 14 TeV will be able to explore scenarios with Wino masses up to around 600 GeV with 300 fb$^{-1}$. Moreover we show that a 100 TeV collider has the potential to probe Wino mass around 3 TeV with 300 fb$^{-1}$, having therefore the possibility to rule out anomaly mediation scenarios with long lived chargino. An increase in the luminosity for a $100$ TeV collider or an increase in the energy would be enough to probe the thermal dark matter mass range: a 200 TeV collider with 1000 fb$^{-1}$ of luminosity would extend the reach to over 3 TeV.

\section{Direct detection interplay}
\label{sec:dd}

\begin{figure}[t]
\centering
\subfigure[Higgs.\label{fig:Higgs}]{\includegraphics[width=0.34\textwidth]{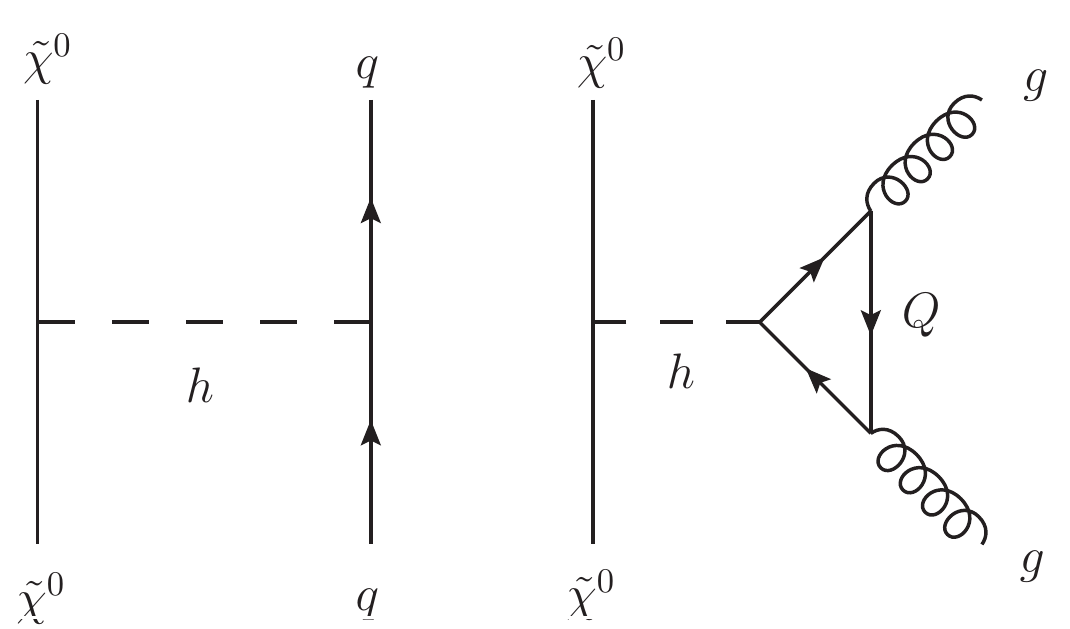}}
\hspace{0.5cm}
\subfigure[Twist-2.\label{fig:twist-2}]{\includegraphics[width=0.14\textwidth]{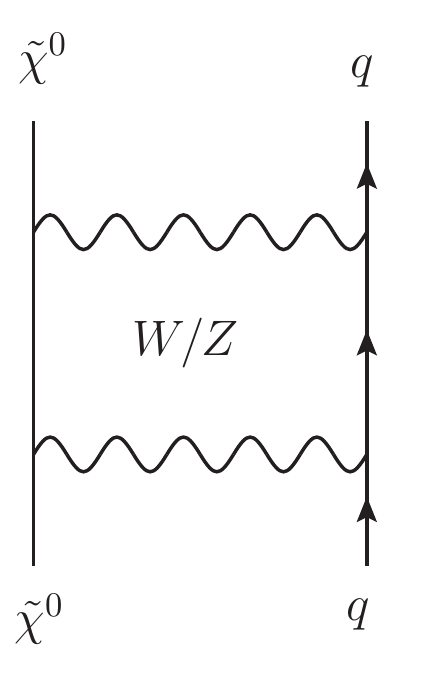}}
\hspace{0.5cm}
\subfigure[Gluon.\label{fig:gluon}]{\includegraphics[width=0.21\textwidth]{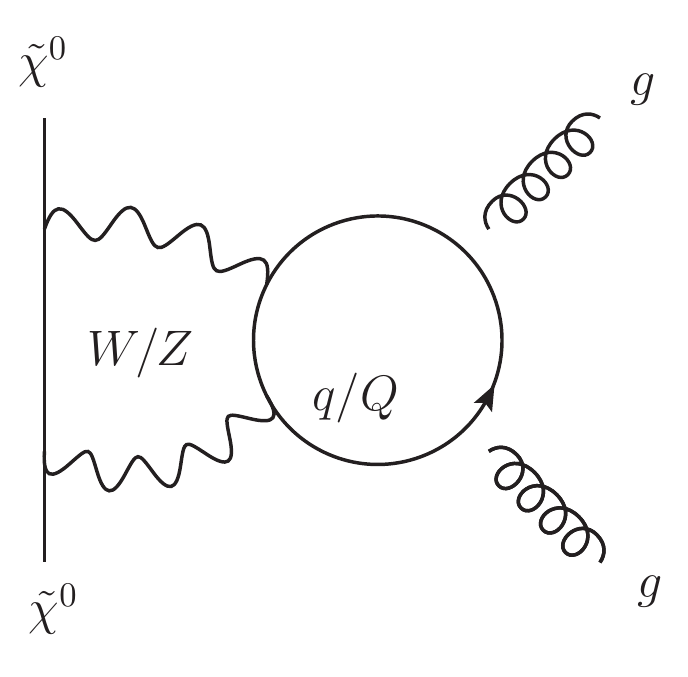}\includegraphics[width=0.21\textwidth]{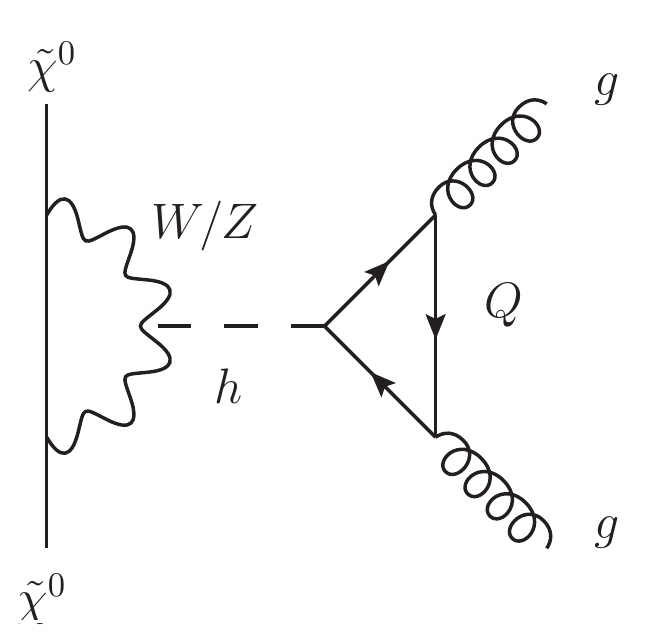}}
\caption{Diagrams that contribute to the neutralino nucleon cross section. The diagrams with $Z$ exchange are relevant only for the Spin Dependent cross section.
\label{fig:Feyn} }
\end{figure}

Direct detection experiments exploit the recoil energy from dark matter particle scattering on nuclei \cite{Goodman:1984dc}. The spin independent cross section of the LSP with a nucleon N can be expressed as
\begin{equation}
\sigma_N^{SI} = |\mathrm{Higgs}+\mathrm{gluon}+\mbox{twist-2}|^2,
\end{equation}
where $Higgs$, $gluon$ and $twist$-$2$ refer to the diagrams of figure \ref{fig:Feyn}. The $twist$-$2$ diagram contributes to the amplitude with opposite sign with respect to the other diagrams and it will then lead to an accidental cancellation. 

\begin{figure} [tb]
\centering%
   \begin{minipage}{0.69\linewidth}
     \includegraphics[scale=0.80] {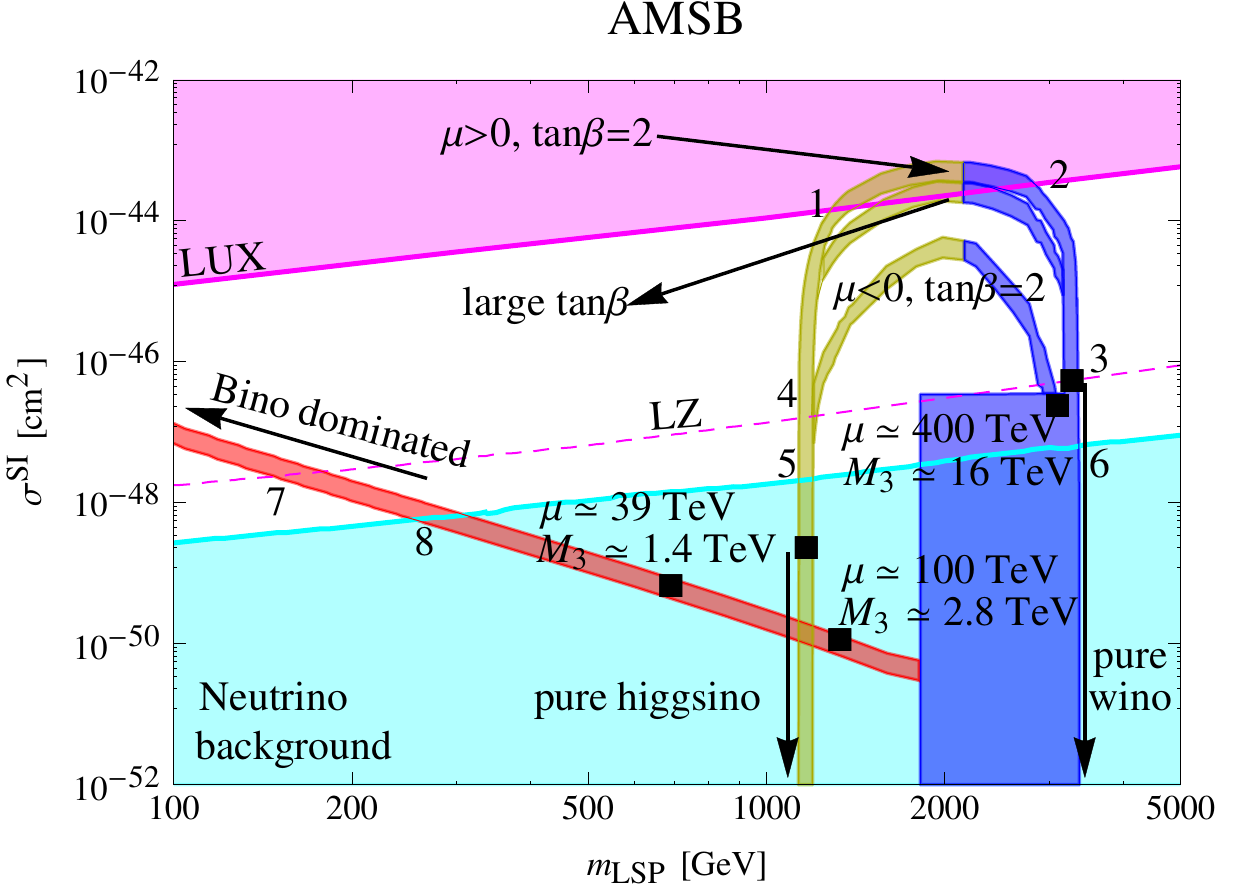}
   \end{minipage}
   \begin {minipage}{0.29\linewidth}
     \includegraphics[scale=0.5] {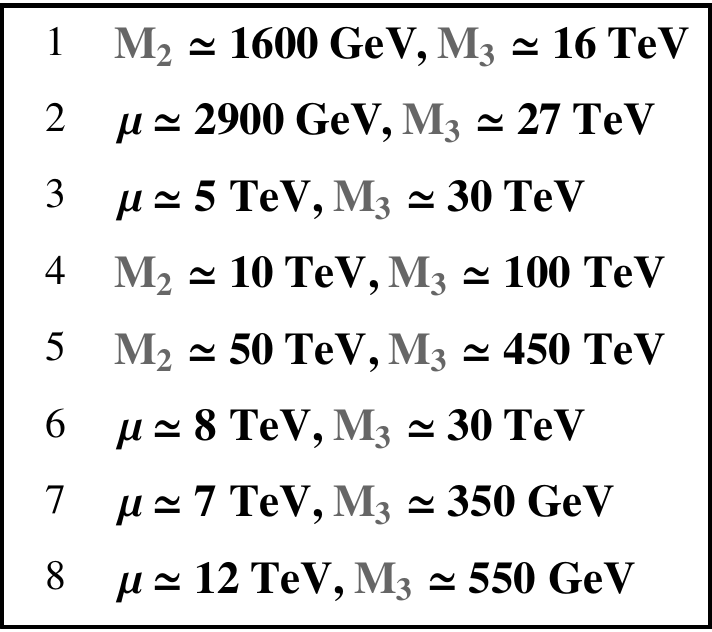}
   \end{minipage}
\caption{Spin independent nucleon-neutralino cross section in a split SUSY with anomaly mediation model (the red curve describe a Bino-like LSP, the yellow curve a higgsino-like and the blue one a Wino-like dark matter candidate). The magenta area is the actual bound by LUX, the dashed magenta line is the projected reach of LZ. The neutrino background is shaded in light blue. The numbers are explained in the legend on the right panel.} 
\label{fig:AMSB}
\end{figure}

The spin independent cross section that satisfy the relic density constraint is shown in figure~\ref{fig:AMSB}. 
The magenta area represents the region excluded by LUX \cite{Akerib:2013tjd}, while the dashed magenta curve shows the projected reach from LZ \cite{LUXLZ,Cushman:2013zza}. The light blue area describes the irreducible neutrino background \cite{Billard:2013qya}. The yellow, blue and red curves represent respectively the cross section for higgsino, Wino and Bino dark matter. We will describe the curve starting from the pure higgsino case and then discuss the behaviour as the higgsino mass is increased. The relic density constraints a pure higgsino dark matter to have a mass around $1.1$ TeV. In this limit the dominant contributions are the $gluon$ and $twist$-$2$ diagrams. However a cancellation suppress the spin independent cross section:
\begin{equation}
\sigma_N^{SI}\lesssim 10^{-48} \mbox{cm}^2.
\end{equation}

While the higgsino mass increases, $M_2$ decreases and the $Higgs$ diagrams start to be of the same order of the other contributions. There is a value of $M_2$ such that the $Higgs$ contribution accidentally cancels the other diagrams. Continuing along the yellow curve, $M_2$ decreases and eventually, when $\mu\sim M_2$, the $Higgs$ diagrams dominate. The LUX bounds apply in the region of maximal mixing. The three different yellow/blue curve show the cross section for negative and positive $\mu$ with small $\tan\beta$ and the limit for large $\tan\beta$. At large $\tan\beta$ the sign of $\mu$ becomes irrelevant. When $\mu<M_2$ the cross section decreases and the LSP approaches the pure Wino state. In the region $\mu\gg M_2$ the $gluon$ and $twist$-$2$ diagrams are also important. As it was for the pure higgsino case, the $gluon$ and $twist$-$2$ contributions accidentally cancel suppressing the cross section: there is a value of $\mu$ such that the $Higgs$ diagrams cancel the $gluon$ and $twist$-$2$ contribution and the cross section vanishes. When the higgsino and the Bino are decoupled the cross section is dominated by the $gluon$ and $twist$-$2$ diagrams:
\begin{equation}
\sigma_N^{SI}\lesssim 10^{-47} \mbox{cm}^2.
\end{equation}
Increasing further the value of $\mu$, $M_1$ decreases and the cross section stays constant, this is the top flat edge of the blue rectangle. Once the Bino becomes the LSP, the $Higgs$ diagrams are the only contributions to the cross section and they are suppressed by the large value of $\mu$. In order to have the correct relic density, the Bino must coannihilate with the Wino and in order to decrease the LSP mass scale and maintain the right splitting for coannihilation, $\mu$ must decrease. When $\mu$ decreases the cross section increases. 

\begin{figure} [tb]
\centering
\begin{center} 
\includegraphics[scale=0.58] {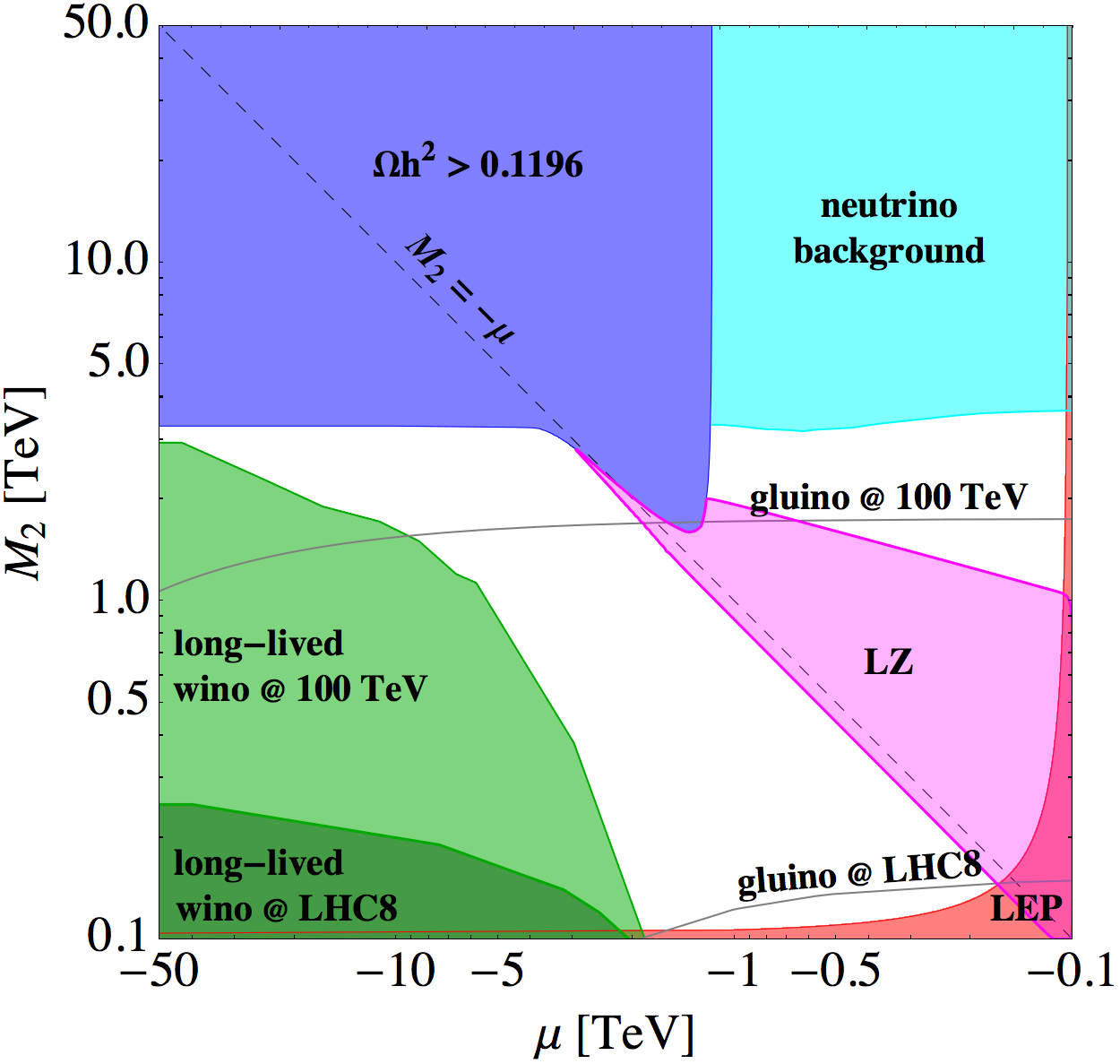}
\includegraphics[scale=0.58] {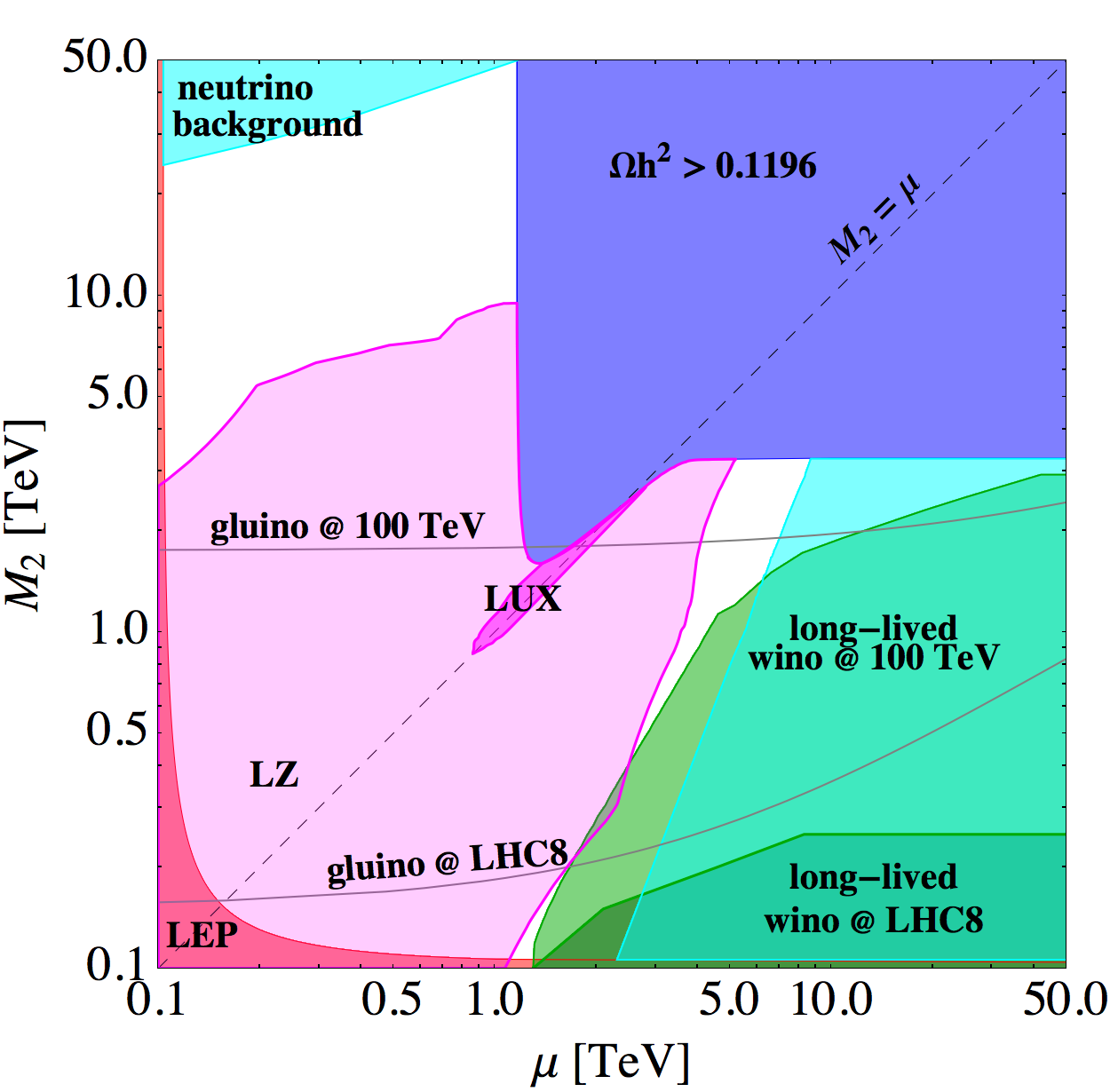}
\includegraphics[scale=0.58] {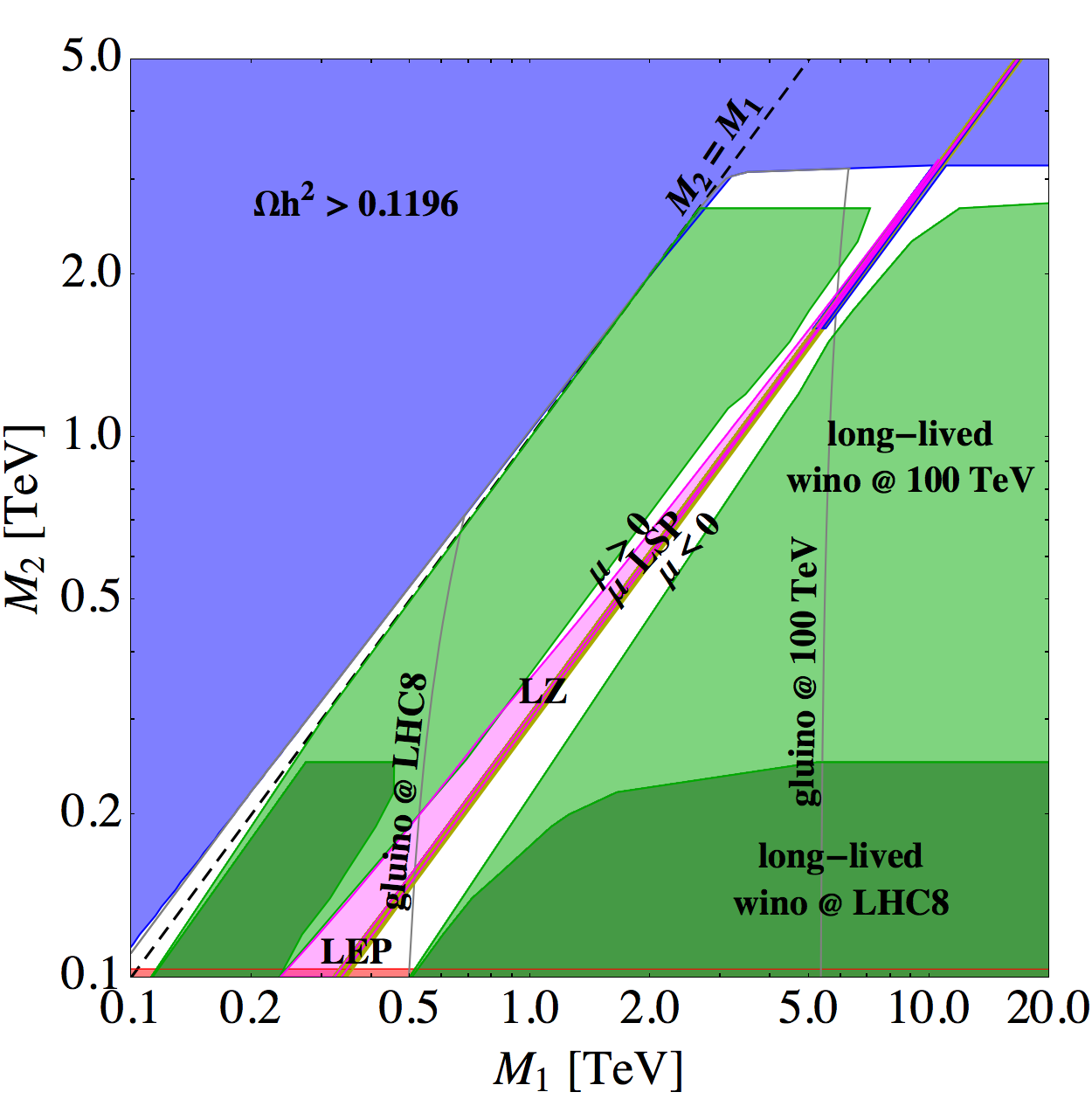}
\end{center}
\caption{The parameter space region allowed by the requirement that the neutralino relic abundance does not exceed the relic density in the plane  ($\mu$, $M_2$) for $\mu<0$ ($\mu>0$ ) in the left (right) panel. The bottom panel shows the plane ($M_1$, $M_2$). Direct Detection, collider constraints and future reach are also shown.} 
\label{fig:relicAMSB} 
\end{figure}

The interplay between collider and direct detection searches is shown in figure \ref{fig:relicAMSB}. In the upper panels there are the $(\mu,M_2)$ planes for both negative (left) and positive (right) $\mu$. The relic density constraint excludes the blue region. The bounds at LHC8 and projected reach at a 100 TeV collider for long-lived Winos searches are shown by the dark and light green area respectively. The area below the grey lines is excluded by LHC8 or it may be probed at a 100 TeV collider through gluino pair searches. In the negative $\mu$ scenario the direct detection searches are limited due to the suppression of the cross section, while for positive $\mu$ direct detection and collider searches are complementary. The bottom panel describes the $(M_1,M_2)$ plane. The yellow line divides the plane in two region with opposite sign of $\mu$ and represents the area in which the higgsino is the LSP. We have to notice how long-lived Wino searches would be able to probe large part of the parameter space, leaving unexplored regions corresponding to the pure Wino and higgsino cases.

\section{Conclusion}

The fact that the LHC excludes electroweakinos up to few hundred GeV, well below the interesting case of SUSY dark matter, motivates studies of mass reach at future hadron colliders.

In this talk, we presented the mass reach for long lived searches in split SUSY models with anomaly mediation at future hadron colliders, the implications for dark matter and the complementarity with direct detection experiments. In the original paper \cite{diCortona:2014yua}, we presented also mass reach for scenarios with Wino NLSPs decaying to leptons (or $b$-jets and leptons) and Bino LSP, with Bino NLSP decaying to photon and gravitino and with higgsino NLSP decaying to gravitino and Z or W bosons in gauge mediation models. Moreover we presented interplay between collider searches and direct detection experiments for models of split SUSY with universal gaugino masses.

In anomaly mediation scenarios, the ratio between the gluino and the Wino mass is large, making Wino searches more powerful. Indeed a 100 TeV collider may be able to reach pure Wino dark matter (and indirectly gluinos heavier than 20 TeV) with 3000 fb$^{-1}$ of luminosity. Long-lived Winos set strong mass reach in regions of the parameter space where direct detection experiments are weak: the interplay between collider and direct detection experiments may be able to cover large areas of the parameter space of neutralino dark matter. 

In conclusion we showed how future hadron colliders can explore the neutralino dark matter parameter space in split SUSY with anomaly mediation, addressing the necessity of colliders beyond the LHC.

\end{document}